\documentclass[sigplan,10pt,screen]{acmart}
\renewcommand\footnotetextcopyrightpermission[1]{}
\usepackage{xspace}
\usepackage{tabularx}
\usepackage{subfig}
\usepackage{enumitem}
\usepackage{algorithm}
\usepackage{algorithmicx}
\usepackage[noend]{algpseudocode}
\usepackage{enumitem}
\usepackage{pifont}
\setitemize{noitemsep,topsep=0pt,parsep=0pt,partopsep=0pt}

\newcommand{\parabf}[1]{\medskip\noindent\textbf{#1}}

\newcommand{\cut}[1]{}

\newcommand{\sysname}{ExpertPlex\xspace}

\begin{document}
\title{\sysname: A High-Goodput Disaggregated Serving System for MoE LLMs with Adaptive Persistent Kernels}
\author{
\vspace{-0.1in}
Bingyang Wu$^{1}$\qquad Chao Jin$^{1}$\qquad Zili Zhang$^{1}$\qquad Xinming Wei$^{1}$\\
Yinmin Zhong$^{1}$\qquad Ruidong Zhu$^{1}$\qquad Chengxu Yang$^{2}$\qquad Xin Jin$^{1}$\qquad Yuliang Liu$^{2}$\\
$^{1}$\textit{Peking \  University} \qquad $^{2}$\textit{Independent Researcher}
\vspace{-0.1in}
}

\begin{abstract}
LLMs scale Mixture-of-Experts (MoE) parameters for superior intelligence, but massive weights and dynamic computation impede efficient serving.
Existing instance-level prefill-decode disaggregation isolates the phases on separate full-model replicas.
As MoE weights grow, each instance may span tens to hundreds of GPUs, making resource allocation increasingly coarse.
Configured prefill-to-decode ratios thus often mismatch demand, overprovisioning one phase while overloading the other.
Prefill-decode colocation avoids this duplication, but existing Green Context solutions partition each GPU by phase and fix phase resources during a kernel.
They cannot track resource changes across operations or layerwise variation in routed expert load, causing head-of-line blocking or idle reserved resources.
Partitioning every GPU also leaves each phase with fewer local resources, forces wider parallelism and more communication, and lets prefill and decode traffic interfere on the shared network.

We present \sysname, which shares massive MoE experts across phases while disaggregating lightweight attention modules.
Expert sharing eliminates over 95\% of duplicate model weights and multiplexes dynamically sparse computation, while attention disaggregation reduces attention communication cost.
\sysname further uses (1) adaptive persistent kernels to schedule dynamic expert computation at tile granularity for efficient, isolated execution;
(2) attention-initiated MoE communication to avoid network interference and enable cross-phase communication-computation overlap; and
(3) a tile-to-cluster model to optimize these mechanisms for maximum goodput.
Experiments serving MiniMax-M2.7 and GLM-5.1-FP8 show that \sysname improves goodput by up to 2.01$\times$ over instance-level prefill-decode disaggregation and 1.66$\times$ over prefill-decode colocation.
\end{abstract}

\maketitle
\pagestyle{plain}
\thispagestyle{plain}


\section{Introduction}
\label{sec:introduction}

Scaling laws continue to reward larger model capacity~\cite{kaplan2020scaling,brown2020language}.
Frontier LLMs therefore increasingly use sparse Mixture-of-Experts (MoE) modules that dynamically activate a few experts to add parameters without proportionally increasing computation~\cite{deepseekv3,minimaxm1,glm45,kimik2,qwen3,mimov2flash}.
It is still challenging to serve them efficiently, because instances must still store massive expert weights, while token-dependent routing makes expert computation and communication vary across tokens.

Modern LLM serving systems typically use prefill-decode disaggregation (PDD) to serve MoE LLMs to avoid interference between the compute-intensive prefill phase and the latency-sensitive decode phase~\cite{zhong2024distserve,patel2023splitwise,hu2024inference,deepseekv3,mooncake}.
For better performance, PDD also needs to allocate resources according to the P:D ratio that matches the resource demands of two phases.
However, existing instance-level PDD provisions a complete model replica for each phase, so its allocation granularity grows with the model.
A reported DeepSeek-V3 deployment uses 32 GPUs for prefill and 320 GPUs for decode in one unit~\cite{deepseekv3} to achieve this ratio.
Small clusters cannot realize this ratio, leading to resource waste.
Large clusters can achieve this ratio, but suffer from a larger failure blast radius because a rank failure in hierarchical communication can stall the entire unit~\cite{deepep,eep}. They can also scale only at the granularity of the large deployment unit when traffic shifts.
Duplicated expert weights also displace KV cache.
PDD therefore obtains isolation at the cost of memory efficiency, elasticity, and fault containment.

Prefill-decode colocation solutions instead use Green Context~\cite{greencontext} to reserve SMs for each phase to achieve weight deduplication and computation multiplexing~\cite{shi2025nexus,muxwise,hong2025semi}, but they cannot follow the dynamic demands of MoE LLMs.
When Green Context is used, resources are fixed for a kernel, and reconfiguration is limited at the prefill layer level because it requires CPU intervention.
However, MoE and attention modules in a layer and communication and computation operations in MoE modules have different demands~\cite{step3afd,megascaleinfer}.
Dynamic expert computation also creates layer-to-layer changes in expert activation on each GPU~\cite{wei2026ultraep,he2021fastmoe,he2022fastermoe}.
Coarse reconfiguration therefore creates head-of-line blocking and resource bubbles (\S\ref{sec:background}).
Partitioning a GPU also reduces the local resources available to each phase, increasing its required degree of parallelism with more communication while leaving cross-phase network interference unmanaged.

To address these problems, our key insight is a hybrid disaggregation-colocation architecture that shares experts across phases but disaggregates their attention modules.
MoE weights contribute over 95\% of model parameters. Sharing eliminates their cross-phase memory duplication and multiplexes dynamic per-rank expert loads, letting computation from either phase fill bubbles in the other's attention--expert pipeline.
Attention holds under 5\% of parameters, so disaggregation enables independent per-phase allocation in single-GPU units without duplicating massive expert weights.
Because attention is more compute-intensive~\cite{megascaleinfer,step3afd},
giving each phase whole attention GPUs rather than intra-GPU partitions preserves local compute capacity for each phase in a GPU. Partitioning attention GPUs would require wider per-phase parallelism for equivalent resources, increase communication, and introduce cross-phase attention network interference.
This architecture lets \sysname match phase demands with fewer GPUs and less traffic, improving utilization, scaling granularity, and fault containment.

However, sharing the experts requires finer GPU control than existing mechanisms provide.
CUDA stream priorities~\cite{cuda} and API interception~\cite{tgs} cannot preempt long-running prefill kernels that may significantly block latency-sensitive decode kernels, while Green Context~\cite{greencontext}, MPS~\cite{mps}, and MIG~\cite{nvidia-mig} cannot reallocate resources at fine granularity between phases, leading to resource bubbles.
Shared experts also couple the phases through communication.
The traffic flows from two phases to the same MoE module can interfere with each other, and conventional two-sided MoE communication requires MoE-side coordination that can cause deadlock when different ranks are used by different phases.

To address these challenges, we propose \sysname, a disaggregated serving system for MoE LLMs with adaptive persistent kernels.
\sysname runs an \textit{Adaptive Persistent Kernel} (APK) on each MoE GPU.
APK schedules MoE computation at tile boundaries, providing bounded preemption for urgent decode work and reallocating idle CTA clusters to prefill without CPU intervention or kernel relaunch.
The bound is independent of sequence length, while persistent execution preserves CUDA Graph compatibility and GPU utilization.
\sysname further uses \textit{attention-initiated one-sided MoE communication} to use mostly disjoint network paths to transfer activations between attention and final MoE-side buffers without APK coordination.
It therefore removes MoE-side polling, avoids cross-phase deadlock and interference, and overlaps communication from one phase with computation from the other.
Because these mechanisms introduce coupled choices in placement, parallelism, computation-communication overlap, and tile scheduling,
\sysname proposes a cross-stack placement optimizer that models them jointly to maximize goodput.
Evaluations serving MiniMax-M2.7 and GLM-5.1-FP8 on real-world workloads show that \sysname improves goodput by up to 2.01$\times$ over instance-level prefill-decode disaggregation, and 1.66$\times$ over Green Context-based prefill-decode colocation.

This paper makes the following contributions.
\begin{itemize}[leftmargin=*]
    \item We introduce a hybrid architecture that eliminates cross-phase MoE weight duplication, multiplexes dynamically sparse expert computation, and disaggregates attention to reduce the per-phase degree of parallelism with less communication while preserving fine-grained isolation.
    \item We design APKs for tile-level preemption and reallocation and attention-initiated one-sided communication to reduce network interference and enable cross-phase overlap between communication and computation.
    \item We jointly optimize them across the stack and evaluate \sysname on two frontier MoE LLMs, demonstrating goodput improvements over state-of-the-art baselines.
\end{itemize}

\section{Background and Motivation}
\label{sec:background}

\subsection{MoE LLM Inference}

An LLM stacks Transformer layers, each with an attention module and a feed-forward network (FFN) module.
Attention mixes information across tokens and materializes key-value tensors as the KV cache, which later iterations reuse.
The FFN instead transforms each token independently.
Frontier MoE LLMs replace conventional FFNs with mixture-of-experts (MoE) modules and scale their expert weights aggressively~\cite{deepseekv3,minimaxm1,kimik2,qwen3,mimov2flash}.
These weights now dominate the model footprint, such as 95\% in DeepSeek-V4-Pro~\cite{xu2026deepseek}, 96\% in GLM-5.1-FP8~\cite{zeng2026glm}, and 98\% in MiniMax-M2.7~\cite{chen2026minimax}.

An MoE module partitions the FFN into many experts.
\textit{Shared experts} process all tokens, while a router dynamically selects a small top-$k$ subset of \textit{routed experts} for each token at each layer.
The router dispatches the token activation to the activated experts and combines their outputs before the next layer.
This sparse activation increases model capacity without proportionally increasing computation per token.

Inference has two phases with different objectives.
\textit{Prefill} processes all input tokens in parallel, builds their KV cache, and produces the first output token, which makes it throughput-oriented.
\textit{Decode} then generates one token per iteration from the preceding token based on the prefix KV cache.
It is latency-sensitive because decode latency is user-visible and orders of magnitude shorter than prefill latency.

\subsection{GPU Execution Model}

\begin{figure}[t]
    \centering
    \includegraphics[width=0.95\linewidth]{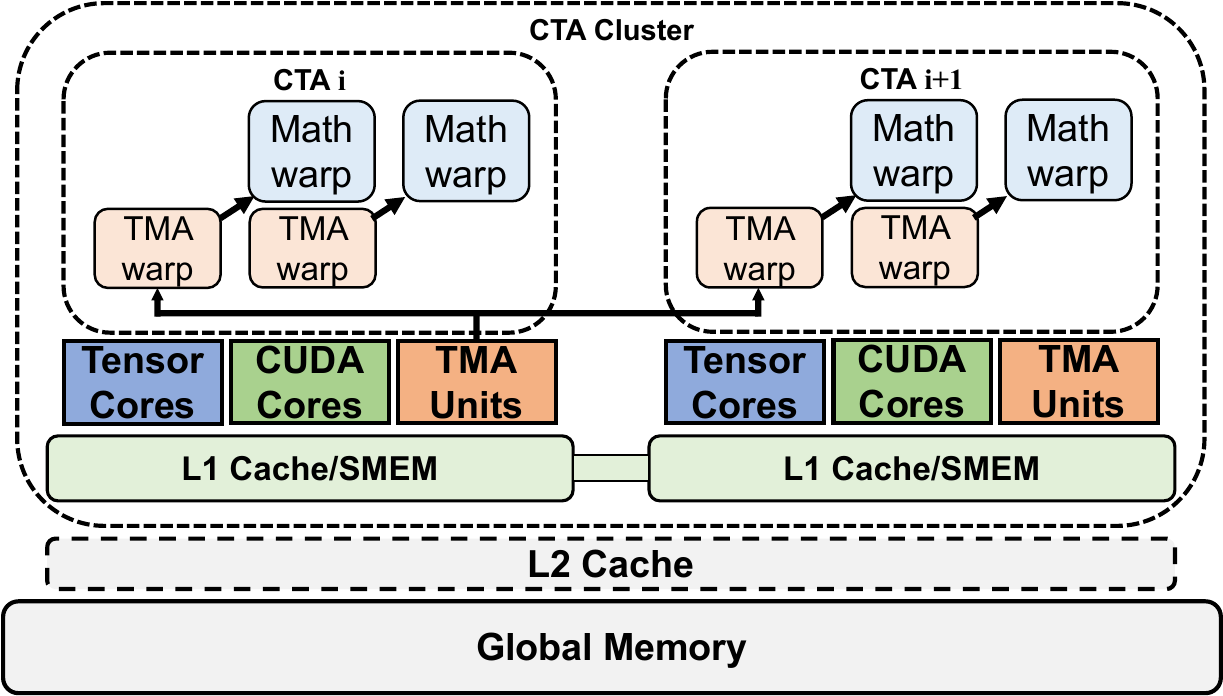}
    \caption{GPU execution model.}
    \label{fig:background:gpu_execution}
    \vspace{-0.2in}
\end{figure}

\begin{figure*}[t]
    \includegraphics[width=\linewidth]{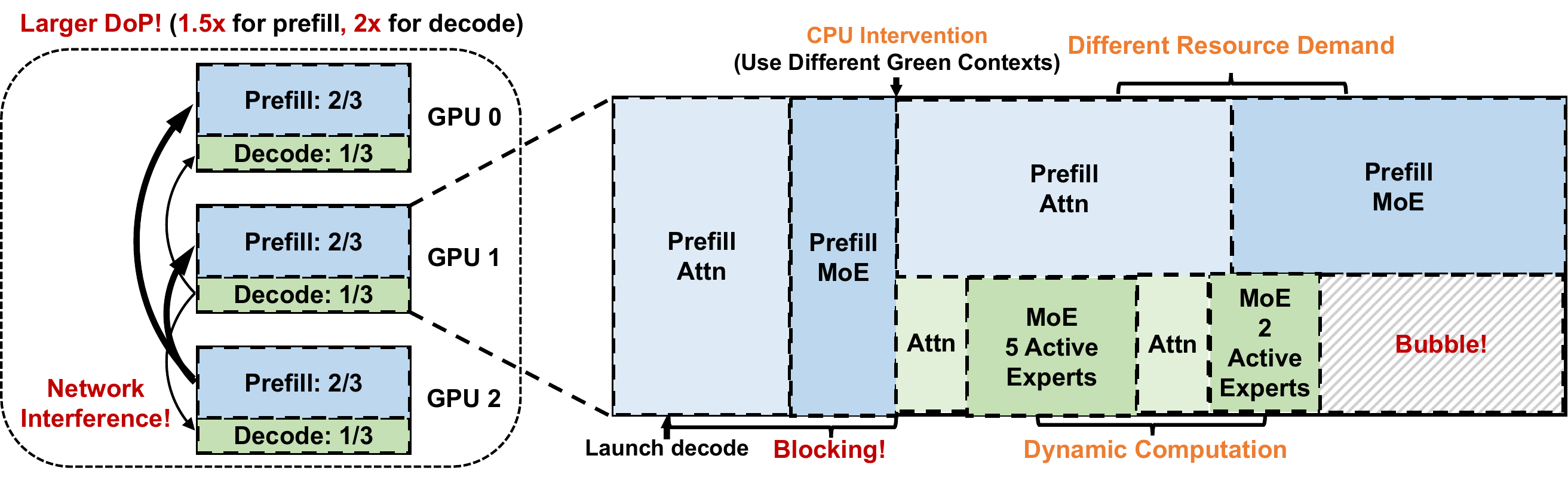}
    \caption{Limitations of prefill-decode colocation solutions.}
    \label{fig:background:pd_colocation}
    \vspace{-0.1in}
\end{figure*}

MoE modules are primarily grouped general matrix multiplications (GEMMs) on GPUs, so their performance depends on the GPU execution model.
As Figure~\ref{fig:background:gpu_execution} shows, a GPU contains many streaming multiprocessors (SMs).
Each SM integrates Tensor Cores, CUDA cores, Tensor Memory Accelerator (TMA) units, registers, an L1 cache, and software-managed shared memory.
All SMs share the L2 cache and global memory, which is also visible to peer GPUs and NICs.
Recent GPUs allow a thread-block cluster to coordinate across SMs through distributed shared memory (DSMEM) and to multicast a global-memory tile through TMA.

GPU work is launched as kernels through CUDA streams.
Kernels in one stream execute in order, while kernels in different streams may overlap when resources are available.
Within a kernel, threads form warps, warps form cooperative thread arrays (CTAs), and CTAs may form clusters.
A CTA runs on one SM until completion, while CTAs in a cluster can cooperate across SMs.
High-performance GEMM kernels specialize warps for data movement and Tensor Core computation, then pipeline them to hide memory latency~\cite{deepep,deepgemm2025}.

\subsection{Large-scale MoE LLM Serving}

MoE weights often exceed one GPU's memory, so serving systems use expert parallelism (EP) to shard experts across GPUs.
Because routing decisions vary by token, \textit{dispatch} sends activations to the ranks hosting the selected experts and \textit{combine} returns their outputs.
Both operations are commonly implemented as all-to-all communication~\cite{deepep}.
Serving systems can hide part of this cost with two-batch overlap (TBO), which overlaps one microbatch's communication with another's computation, or single-batch overlap (SBO), which overlaps communication with shared-expert computation in the same microbatch.

Attention admits a different parallelization because its weights are much smaller.
Data parallelism (DP) replicates attention to favor throughput and KV-cache capacity with little communication.
Tensor parallelism (TP) shards attention to reduce per-request latency, but adds communication and may reduce effective KV-cache capacity when entries are replicated.
Therefore, the resource demand and best configurations of attention and MoE often differ~\cite{megascaleinfer,step3afd}.

\subsection{Prefill-Decode Disaggregation}

Prefill-decode disaggregation (PDD) places the two phases on different GPU instances to eliminate cross-phase interference and specialize each instance for its phase~\cite{zhong2024distserve,patel2023splitwise,hu2024inference,mooncake}.
Prefill benefits from hierarchical dispatch and combine due to the large number of activations~\cite{deepep}.
An activation crosses the scale-out network once to a remote node and is then multicast over its faster scale-up fabric, avoiding redundant scale-out transfers when several activated experts reside on that node.
Decode moves far less data per iteration, so direct transfers to the relevant ranks reduce latency.
PDD also specializes expert computation~\cite{deepgemm2025}.
Prefill packs many expert inputs into a contiguous layout to reduce memory.
Decode has too few tokens to amortize this packing, so a masked layout avoids copies.

\parabf{Limitations.}
The isolation, however, creates a rigid and increasingly coarse deployment granularity.
Each prefill or decode instance must hold a complete model replica, so growing MoE weights directly increase its minimum GPU count, and duplicated weights displace KV-cache capacity.
PDD must then provision these indivisible instances in a prefill-to-decode resource ratio that matches the phases' asymmetric demands.
The smallest combination that realizes this ratio defines the deployment unit, which becomes much larger than either phase's instance alone.
One reported DeepSeek-V3 unit combines 32 prefill and 320 decode GPUs to realize its target ratio~\cite{deepseekv3}.
Another uses 176 GPUs as a single unit for a different load~\cite{deepseekv3-day6}.
Kimi-K2 has likewise been deployed on 128 H200 GPUs~\cite{sglang-kimik2-ep}.

This compounding effect makes the target ratio impossible to realize in a small cluster, while deviating from it makes one phase overloaded and leaves the other phase overprovisioned.
For a large cluster with sufficient GPUs, each scaling step changes capacity by hundreds of GPUs and cannot match moderate traffic shifts efficiently.
Larger units also widen the failure blast radius.
Unlike direct transfers, hierarchical communication couples many ranks into one group, so a single rank failure can stall the entire unit and is difficult to recover from~\cite{eep,deepep}.
PDD therefore obtains isolation at the cost of memory efficiency, elasticity, and fault containment.

\subsection{Prefill-Decode Colocation}

Colocation avoids separate model replicas by sharing one instance across phases.
To avoid interference, many techniques are proposed.
Chunked prefill splits a long input into chunks that can be mixed with decode iterations to mitigate interference~\cite{sarathi,podattention}.
However, it introduces memory access overhead from rereading prefix cache entries and model weights~\cite{zhong2024distserve}.
Recent systems colocate the phases through spatial GPU partitioning~\cite{muxwise,shi2025nexus,hong2025semi}.
They commonly use Green Contexts~\cite{greencontext} to reserve separate SMs for prefill and decode, avoiding duplicate model replicas while providing compute isolation.

\parabf{Limitations.}
However, spatial partitioning cannot handle temporal resource demand variations along three dimensions.
First, under EP, the number of activated experts and the number of tokens on a GPU change across layers, so the MoE load varies by rank and layer~\cite{wei2026ultraep,he2021fastmoe,he2022fastermoe}.
Second, attention and MoE have different resource demands even within the same layer~\cite{step3afd,megascaleinfer}.
Third, each MoE module alternates among dispatch, expert computation, and combine, shifting its bottleneck between communication and computation~\cite{liu2026revealing}.
A fixed partition cannot follow these changes.
Changing an allocation requires CPU coordination, and kernel-completion waits make per-kernel changes impractical.
Existing systems therefore repartition at the prefill layer boundary~\cite{muxwise,shi2025nexus}.

As Figure~\ref{fig:background:pd_colocation} illustrates, a poor partition creates two failure modes.
If prefill holds too many SMs when decode becomes ready, latency-sensitive decode waits behind a non-preemptible prefill kernel.
This head-of-line blocking can be severe because a prefill kernel may run for tens to hundreds of milliseconds while a decode kernel finishes in hundreds of microseconds, a difference of orders of magnitude.
If resources remain reserved for decode when it has no ready work, prefill cannot use them, and the GPU develops a resource bubble.
Rank-varying EP load and different module characteristics compound both problems.

Existing solutions also leave communication unisolated.
Prefill and decode dispatch and combine share the same NICs and links, where a large prefill transfer can throttle a small latency-sensitive decode transfer.
Conventional two-sided communication further requires matching two-sided progress.
Independent scheduling can leave some MoE ranks serving prefill while others serve decode, so each phase waits for receiver work on ranks occupied by the other, and the system deadlocks.
Finally, partitioning every GPU gives each phase fewer local resources.
As Figure~\ref{fig:background:pd_colocation} shows, meeting the same latency target may then require a larger degree of parallelism, which increases communication and further amplifies network interference.

\section{\sysname Overview}
\label{sec:overview}

\begin{figure}[t]
    \centering
    \includegraphics[width=0.95\linewidth]{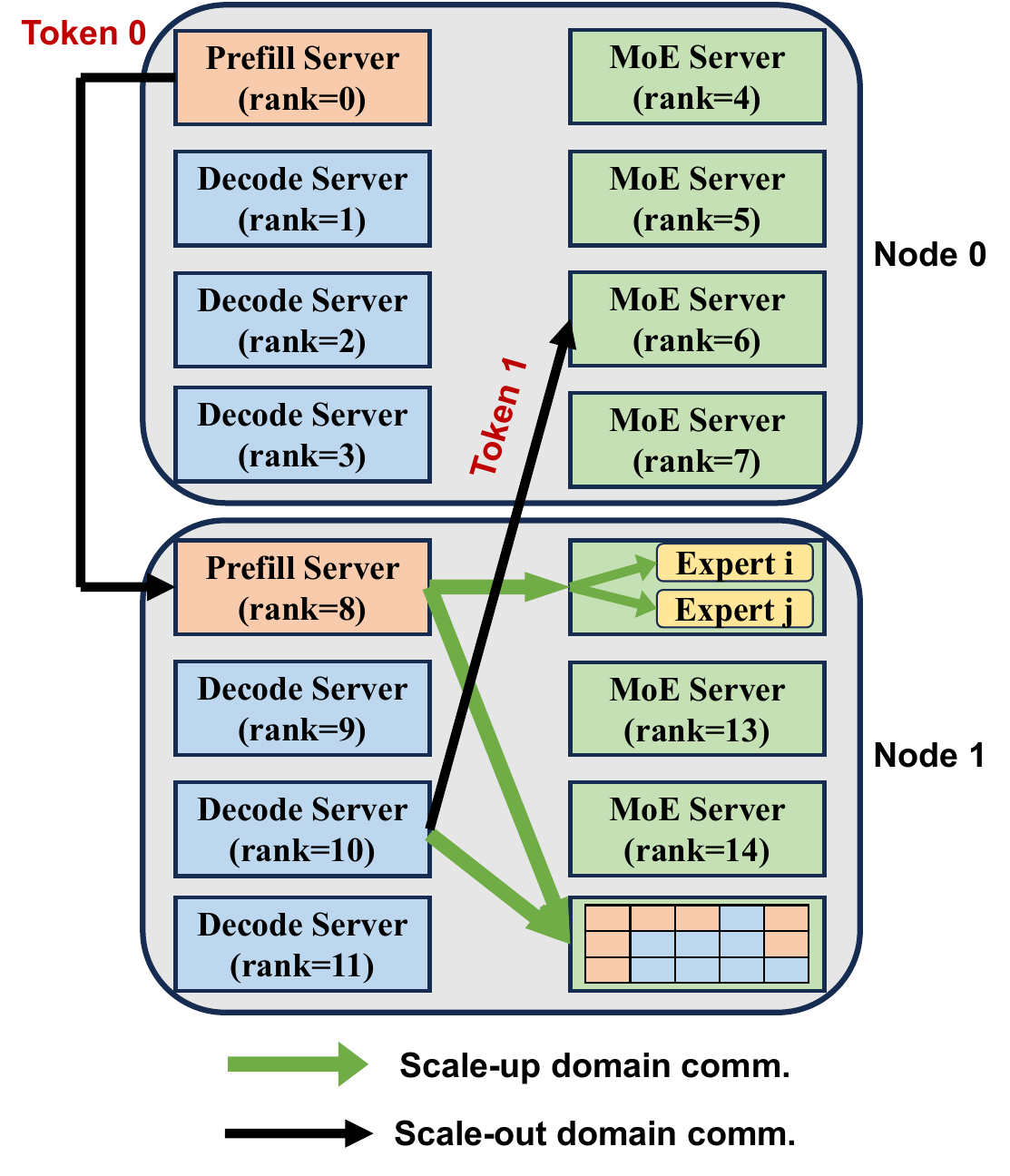}
    \vspace{-0.2in}
    \caption{Architecture of \sysname.}
    \label{fig:overview:architecture}
    \vspace{-0.3in}
\end{figure}

\begin{table*}[t]
\begin{center}
    \begin{tabularx}{\textwidth}{lXXXXXXl}
    \toprule
        & \centering\textbf{API Interception}~\cite{tgs,preemptionInference,gpreempt}
        & \centering\textbf{CUDA streams}~\cite{cuda}
        & \centering\textbf{NVIDIA MPS}~\cite{mps}
        & \centering\textbf{Green Context}~\cite{greencontext}
        & \centering\textbf{NVIDIA MIG}~\cite{nvidia-mig}
        & \centering\textbf{\sysname}
        &
        \\
    \midrule
    \textbf{CUDA Graph Compatibility}
        &
        & \centering$\checkmark$
        & \centering$\checkmark$
        & \centering$\checkmark$
        & \centering$\checkmark$
        & \centering$\checkmark$
        &
        \\
    \textbf{Temporal Multiplexing}
        & \centering$\checkmark$
        & \centering$\checkmark$
        &
        &
        &
        & \centering$\checkmark$
        &
        \\
    \textbf{Spatial Multiplexing}
        & 
        & 
        & \centering$\checkmark$
        & \centering$\checkmark$
        & Limited
        & \centering$\checkmark$
        &
        \\
    \textbf{Bounded Fast Preemption}
        &
        &
        &
        &
        &
        & \centering$\checkmark$
        &
        \\
    \textbf{Bounded Fast Reallocation}
        &
        &
        &
        &
        &
        & \centering$\checkmark$
        &
        \\
    \bottomrule
    \end{tabularx}
    \vspace{-0.05in}
    \caption{Comparison of \sysname with existing GPU sharing solutions.}
    \label{system_properties}
    \vspace{-0.3in}
\end{center}
\end{table*}

To address these problems, we propose \sysname, a disaggregated serving system for MoE LLMs with adaptive persistent kernels.
\sysname shares MoE experts across phases while disaggregating attention by phase.
As shown in Figure~\ref{fig:overview:architecture}, each node assigns a configurable subset of GPUs to prefill servers, another subset to decode servers, and the remaining GPUs to MoE servers. Prefill and decode servers in each node can be empty.
Prefill and decode servers run their respective attention modules, while MoE servers host the MoE experts and execute expert computation for both phases.
Because attention weights constitute only a small fraction of the model parameters while MoE weights account for over 95\% of them, sharing MoE servers eliminates massive MoE weight duplication and multiplexes dynamically sparse expert computation.
Disaggregating attention gives each phase full GPUs instead of GPU partitions, reducing its required degree of parallelism and communication while eliminating cross-phase attention network interference.
This MoE-weight-independent boundary matches phase demands with fewer GPUs and enables finer-grained elastic scaling with a smaller failure blast radius.

Each MoE server runs an \textit{Adaptive Persistent Kernel} (APK) that keeps the GPU busy with adaptive scheduling (\S\ref{sec:design:apk}).
APK preserves CUDA Graph compatibility and supports spatial and temporal multiplexing to maximize computation efficiency.
Its tile-level scheduler follows real-time expert load, preempts prefill for decode within a sequence-length-independent bound, and reallocates idle SMs without CPU intervention while preserving prefill performance.

\sysname uses attention-initiated one-sided primitives for MoE communication (\S\ref{sec:design:comm}).
Attention servers push dispatch data into and pull combine results from final MoE-side buffers.
Direct final-buffer access removes MoE-side coordination and polling, avoids deadlock, and overlaps communication with computation across phases.
To reduce network interference, prefill scale-out traffic travels between attention servers whenever possible, while decode communicates directly between attention and MoE servers.

Finally, a cross-stack optimizer models \sysname from the tile level to the cluster level (\S\ref{sec:design:placement}).
It jointly selects APK policies, placements, resource ratios, parallelism, and overlap strategies to maximize goodput under both phases' SLOs.

\vspace{-0.15in}
\section{Adaptive Persistent Kernel}
\label{sec:design:apk}

\subsection{GPU Sharing Design Space}

Across MoE LLMs, prefill processes all input tokens, whereas each decode iteration processes only one new token per request.
The duration gap between their operations therefore grows with input length.
For example, with EP4 on MiniMax-M2.7, grouped GEMMs for decode with eight active experts take 17.7--34.7~$\mu$s, while matched prefill GEMMs with 16K tokens take 1.8--2.9~ms, or 84--101$\times$ longer.
The exact ratio is model dependent, but this trend is universal.
Because every layer invokes an MoE operation, blocking can recur at every layer and accumulate into end-to-end decode latency.
Static partitioning avoids this interference only by leaving resources idle whenever a phase is communicating or has no ready operation.
A useful sharing mechanism must prevent repeated blocking without stranding this idle capacity.

Efficient sharing therefore requires five complementary properties, summarized in Table~\ref{system_properties}.
CUDA Graph compatibility avoids per-operation CPU intervention and its launch and synchronization overhead on short decode paths.
Spatial multiplexing lets different SM subsets execute both phases concurrently.
It lends SMs to the computing phase while the other communicates and changes the split across layers as routing shifts the expert load on each rank.
Time multiplexing lets the same SM switch phases over time when its current phase has no ready operation.
Bounded preemption prevents a long prefill operation from causing severe head-of-line blocking, while bounded reallocation lets a newly ready phase reclaim SMs quickly enough to avoid bubbles.

Launch-time mechanisms miss several properties.
API interception~\cite{tgs,gpreempt,preemptionInference} offers only launch-boundary time multiplexing.
It lacks CUDA Graph compatibility, spatial multiplexing, and bounded preemption and reallocation.
CUDA streams with priorities~\cite{cuda} support graph replay and launch-time ordering, but provide no spatial isolation or bounded resource handoff once a kernel begins execution.

Hardware partitioning provides spatial isolation but remains static.
MPS~\cite{mps} and Green Context~\cite{greencontext} support graph-compatible spatial multiplexing, but fix the allocation during a kernel.
They lack in-kernel time multiplexing and bounded preemption and reallocation.
MPS also leaves cache and shared-memory interference on a shared SM unpredictable.
MIG~\cite{nvidia-mig} provides hard but highly restricted spatial isolation.
H100 exposes only 1g, 2g, 3g, 4g, and 7g profiles, so the only two-way split that lets two phases collectively use the GPU's full compute capacity is 3g--4g.
MIG lacks time multiplexing and preemption, and driver-level reconfiguration cannot bound reallocation latency.
It therefore cannot follow the layer-level load changes described above for either phase.

\subsection{Tile-Level Scheduling}

These requirements favor a scheduling unit that is short and native to optimized GPU operations.
APK chooses a tile, the smallest independently completable unit of these operations.
Registers, shared memory, TMA buffers, and Tensor Core shapes bound its dimensions, so longer inputs add tiles without enlarging them.
For example, our SM90 DeepGEMM configuration~\cite{deepgemm2025} uses CTA tiles of at most a 128$\times$192 output block with a 128-wide reduction tile.
Across MoE-side operations, tile boundaries occur every 2.2--25.3~$\mu$s, independent of the total operation length (Figure~\ref{fig:evaluation:preemption_interval}).
Input length stretches an operation by adding scheduling units rather than lengthening each scheduling interval.

APK switches phases only after the current tile commits.
No accumulator, TMA transaction, shared-memory buffer, or communication state remains live at this boundary, so switching needs no checkpoint, restore, or recomputation.
Finer preemption would preserve pipeline state, while kernel- or layer-level scheduling would retain long blocking intervals.
APK keeps each operation's native CTA or cluster shape, TMA multicast, and warp-specialized pipeline.
The scheduler changes only the next tile, leaving each operation's pipeline and synchronization intact throughout execution.

Before APK starts, \sysname builds an operation template including buffers and TMA descriptors for every MoE operation.
At runtime, an attention server signals a ready operation.
CTA 0 instantiates its template into the decode or prefill queue and updates runtime fields such as the layer, token count, and SM budget.
APK schedules at CTA-cluster granularity because CTAs connected by TMA multicast must execute the same tile.
The cluster leader selects a tile and broadcasts its metadata to peer CTAs.
Different clusters execute different phases for spatial multiplexing, while each cluster switches phases at tile boundaries for time multiplexing and fast reallocation.
Because APK is long-lived and all scheduling occurs on the GPU, this process requires neither CPU intervention nor kernel relaunch and remains compatible with CUDA Graph replay throughout execution.

\vspace{-0.06in}
\subsection{Bounded Tile-level Preemption}

\begin{figure}[t]
    \centering
    \includegraphics[width=0.85\linewidth]{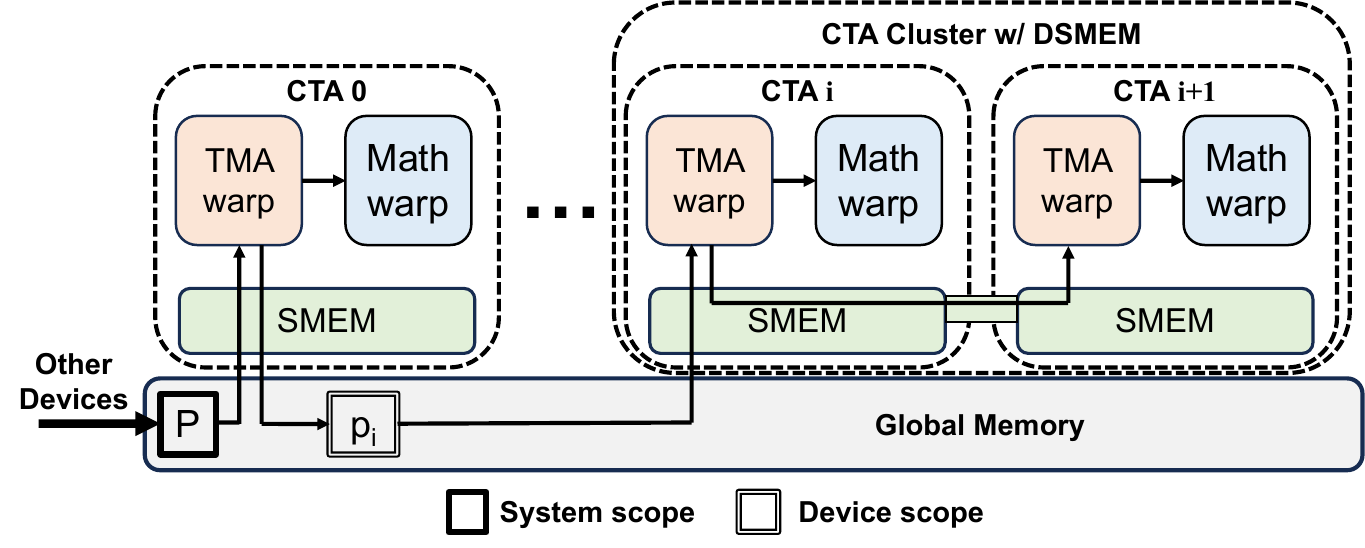}
    \caption{Tile-level preemption mechanism.}
    \label{fig:design:preemption}
    \vspace{-0.15in}
\end{figure}

Independent tile-boundary checks are unsafe because high-performance operations pipeline both warps and CTAs across tiles.
Within a CTA, a TMA warp may load tile $k+1$ while math warps consume tile $k$.
If one warp switches phases, another can wait forever for data or an \texttt{mbarrier} event from the old phase.
Across a CTA cluster, an independently switching CTA can similarly strand peers at a TMA multicast or cluster barrier.
Stalling the entire cluster after every tile prevents both deadlocks but serializes a microsecond-scale pipeline and defeats fine-grained scheduling.

APK instead propagates one cooperative decision through the memory hierarchy shown in Figure~\ref{fig:design:preemption}.
An attention server signals an urgent decode operation through the system-scope word $P$.
CTA 0 checks $P$ at a tile boundary of its current operation and writes one device-scope word $p_i$ for each cluster $i$.
Each cluster reads only its $p_i$, avoiding repeated system-scope accesses and a grid-wide barrier.
At its next tile boundary, the leader CTA broadcasts the decision through DSMEM so every CTA participating in TMA multicast stays on the same operation.
Within each CTA, the first pipeline warp reads the decision, broadcasts it within the warp, and uses an \texttt{mbarrier} handoff to notify later warps before they claim another tile.
The decision remains fixed within a check epoch, so no CTA can claim another old-phase tile after observing it.
The slowest CTA therefore finishes at most its current tile before the cluster converges on the new phase.
The propagation follows the GPU memory hierarchy through system memory, device memory, DSMEM, shared memory, and warp broadcast.
It prevents mixed-phase dependencies both within a CTA and across a CTA cluster without draining the pipeline after every tile.
Preemption is bounded by one tile execution time plus one local cluster check epoch.
This bound remains independent of the interrupted operation's total length.

\section{Attention-initiated MoE Communication}
\label{sec:design:comm}

\subsection{Concurrent Communication over a Shared Network}

Disaggregating prefill and decode attention onto different servers eliminates cross-phase interference for attention communication.
The remaining contention point is therefore MoE communication, because both phases dispatch routed activations to, and combine expert outputs from, the same MoE server pool.
Existing MoE communication usually uses a two-sided protocol for two reasons.
Token routing reveals destinations and transfer sizes only at runtime, and the communication operator is decoupled from the compute operator that consumes its output.
The sender therefore streams data into a bounded receiver-side ring buffer, while the receiver polls and drains that buffer, scatters arrivals into the final tensor, and returns credits that prevent the sender from overwriting unread slots.
The ring buffer bounds memory use, but progress now requires both sides.
With concurrent prefill and decode, independent GPU scheduling can leave some MoE ranks running prefill kernels and others running decode kernels.
Each phase then waits for receiver kernels on ranks occupied by the other phase, so neither ring buffer is drained nor returns the credits needed for its senders to finish.
The blocked sender cannot retire and release its GPU resources for the other phase's receiver kernel.
This cross-phase cycle can deadlock an otherwise correct protocol.

Reserving receiver-side polling SMs avoids this liveness failure only by wasting compute capacity.
MoE traffic is bursty, so these SMs spin between dispatches and cannot execute ready expert tiles.
Letting compute borrow them is unsafe because receiver progress would again depend on whether enough SMs become available at every rank.
Network contention remains as well.
Optimized MoE transfers can saturate the fabric~\cite{deepep}, while scale-out bandwidth is scarce.
For example, DeepSeek-V3 reports 160~GB/s intra-node NVLink bandwidth but only 50~GB/s cross-node InfiniBand bandwidth on H800, a 3.2$\times$ bandwidth gap~\cite{deepseekv3}.
A prefill burst can delay latency-sensitive decode even when the phases use different GPUs.

\subsection{One-Sided Dispatch and Combine}

Our key insight is to remove MoE-side progress with one-sided transfers initiated by attention servers.
\sysname uses both push and pull primitives of the scale-up links, i.e., NVLink, and scale-out links, i.e., RDMA.
APK already preallocates each operation's buffers for its maximum routed-token volume.
This requirement follows from APK's persistent execution because its operation templates, tensor addresses, and communication descriptors must be fixed before runtime scheduling.
\sysname exposes these final dispatch and combine buffers to attention servers and removes the intermediate ring buffer.
Transfers consequently need neither receiver-side draining and scattering nor credit feedback to throttle a sender.
MoE servers expose only buffers and readiness words, with no matching communication kernel or reserved polling SMs.
Dispatch uses an attention-side push.
After routing determines each expert's destination and final offset, the attention server writes activations into that location through NVLink peer stores or one-sided RDMA writes.
It publishes a ready signal only after the payload is visible.
APK does not coordinate this transfer and schedules the MoE task after observing that signal at a tile boundary.

\begin{figure}[t]
    \centering
    \includegraphics[width=0.95\linewidth]{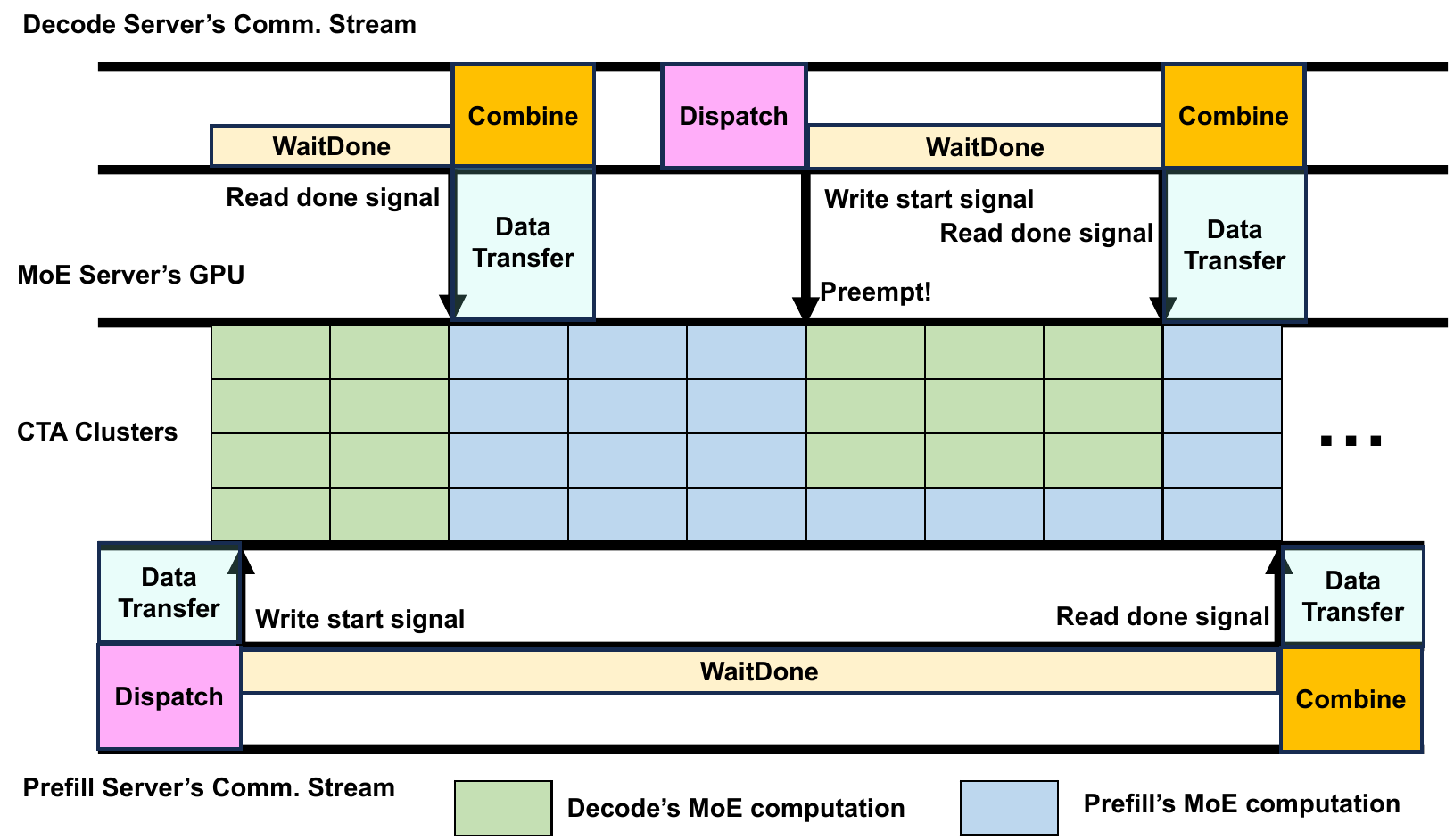}
    \caption{Cross-phase overlap enabled by attention-initiated one-sided transfers.}
    \label{fig:design:comm_overlap}
    \vspace{-0.2in}
\end{figure}

Combine reverses the data flow but remains attention-initiated.
After expert computation writes the final combine buffer and publishes a done signal, a small \texttt{WaitDone} kernel on the attention GPU observes completion, and the attention server pulls the result through NVLink loads or one-sided RDMA reads.
\texttt{WaitDone} is a single-thread kernel, so it can coexist with TBO or SBO and preserve their attention-side overlap.
It runs on the communication stream and blocks only the dependent pull, while the compute stream continues the work exposed by the selected overlap strategy.
Meanwhile, removing MoE-side communication kernels lets APK overlap one phase's MoE computation with the other phase's dispatch or combine, as Figure~\ref{fig:design:comm_overlap} shows.
For example, APK can execute prefill tiles while decode waits to combine, or execute urgent decode tiles while a prefill transfer is in flight.
This mechanism is compatible with each phase's original overlap strategy but adds cross-phase opportunities.
Prefill and decode follow independent dependency chains, so this overlap is not limited by the intra-phase dependencies that constrain TBO and SBO. It can keep the SMs busy as long as there is work available.

\subsection{Hierarchical Prefill Path and Traffic Isolation}

One-sided transfers remove MoE-side coordination, but prefill and decode can still contend for scarce RDMA bandwidth.
Our key insight is to route most prefill scale-out traffic between prefill attention servers, reserving direct attention-to-MoE paths primarily for decode.
Latency-sensitive decode accesses each MoE server directly through local NVLink or remote RDMA.
For a remote node with a prefill server, the source sends each activation once over RDMA to the prefill server with the same local rank on that node.
The receiver multicasts it out over NVLink to all activated experts chunk by chunk.
For combine, the same server gathers outputs over NVLink and returns the combined result through the prefill-only RDMA path.
This path preserves hierarchical deduplication because activations for multiple experts on one remote node cross the scarce scale-out link only once.
The placement optimizer spreads prefill servers across nodes to make this path common (\S\ref{sec:design:placement}).
When a destination node lacks a prefill server, \sysname uses direct one-sided RDMA but assigns prefill a lower-priority InfiniBand virtual lane than decode.
Topology thus separates the phases whenever possible, while virtual-lane priority protects latency-sensitive decode when they must share a scale-out link.

\section{Cross-stack Placement Optimizer}
\label{sec:design:placement}

\subsection{Cross-stack Scheduling Problem}

Although the mechanisms above expose enough control to adaptively schedule prefill and decode with stricter isolation at finer granularity, they also couple decisions conventionally made by different layers of the serving stack.
For example, changing the number of prefill servers simultaneously changes the degree of parallelism, the KV-cache capacity, and the communication path.
Optimizing placement, overlap, or GPU sharing in isolation can therefore select a locally efficient configuration that violates an SLO once these effects interact.
We instead jointly search the attention and MoE parallelism, server placement, overlap strategy, and APK sharing policy.

We formalize a configuration as a layout $\ell$ and a decode SM budget $q$ under contention.
For $(\ell,q)$, let $B_p$ and $B_d$ denote the largest prefill and decode batches whose modeled iteration latencies $T_p$ and $T_d$ meet their respective SLOs, and let $\bar{O}$ be the average output length.
The request-level goodput is defined as
\begin{equation}
G(\ell,q)=\min \left( \frac{B_p}{T_p}, \frac{B_d}{T_d\bar{O}} \right).
\end{equation}
The minimum captures pipeline balance because each request consumes one prefill iteration but $\bar{O}$ decode iterations.
We admit only candidates satisfying both phase SLOs and maximize $G$ over the feasible candidates, rather than overprovisioning either phase in isolation.

\subsection{Tile-aware Latency Model}
\label{sec:design:model}

Evaluating every candidate on a full cluster would make the search prohibitively expensive.
\sysname instead profiles attention and MoE kernels, including APK overhead, on only a few representative GPUs and interpolates their per-rank latency.
Network time requires no profiling GPU. The model computes it directly from routed bytes, measured RDMA or NVLink bandwidth, and the hierarchical path in \S\ref{sec:design:comm}.
For each layout, it derives the work on every rank from the attention and MoE parallelism, placement, routing, and sharing policy.

\begin{figure}[t]
    \centering
    \includegraphics[width=0.95\linewidth]{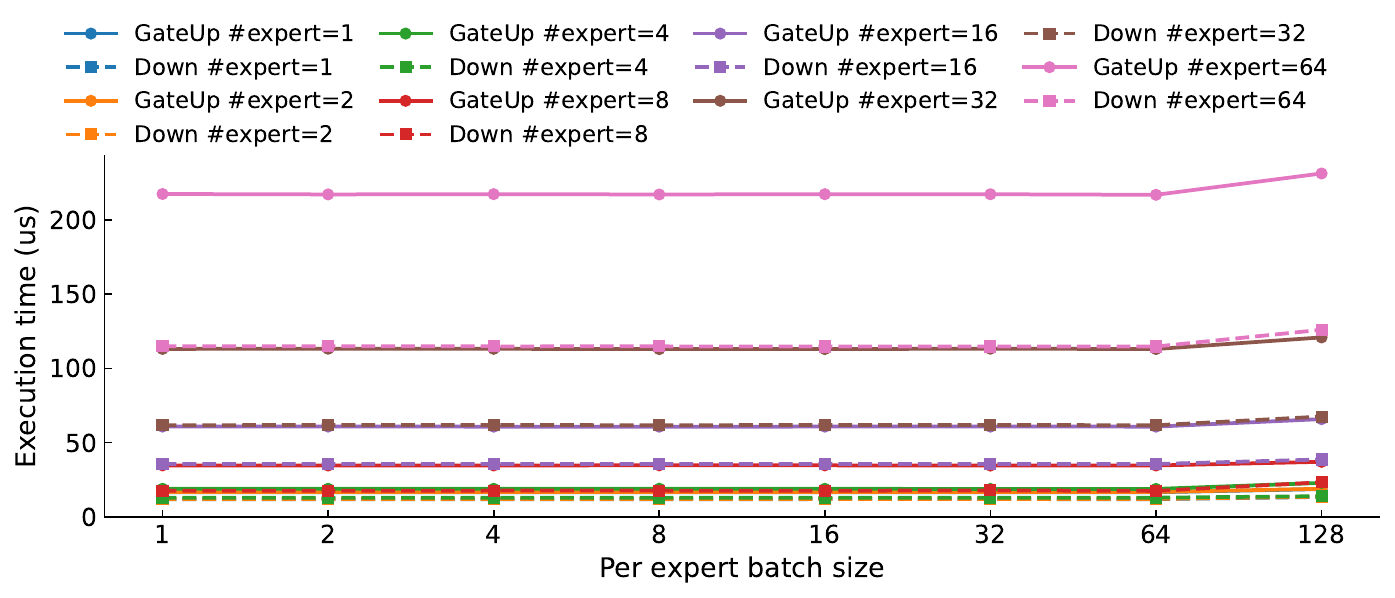}
    \caption{Latency of MoE Grouped GEMMs under different numbers of active experts and tokens.}
    \label{fig:design:minimax_m2_ep4_ordinary_active_experts}
    \vspace{-0.15in}
\end{figure}

For a profiled component $c$, we fit a small set of input samples with
\begin{equation}
\hat{t}_c(x,s)=\alpha_c+\beta_c x+\gamma_c xs+\delta_c xs^2.
\end{equation}
For attention, $x$ is the local batch size and $s$ is the sequence length.
Our key observation is that raw routed-token count is insufficient for MoE computation because its latency follows the number of executed tiles.
If expert $e$ receives $m_e$ token rows and the kernel tile height is $M_t$, the MoE footprint is
\begin{equation}
x_{\mathrm{moe}}=\sum_{e \mid m_e>0}\left\lceil \frac{m_e}{M_t}\right\rceil
\end{equation}
and we evaluate the fitted form with $x=x_{\mathrm{moe}}$ and $s=1$.
Figure~\ref{fig:design:minimax_m2_ep4_ordinary_active_experts} illustrates this distinction. Batches with the same number of tokens can have different latency when they activate different numbers of experts.
This behavior is fundamental to tiled MoE kernels rather than specific to one model or GPU.
Tensor Core and TMA pipelines operate on aligned tiles, and each active expert triggers at least one tile's metadata, weight transfer, and computation even if it receives only a few tokens.
Thus sparse routing changes latency through both token volume and its per-expert packing.
By predicting the per-expert token vector and interpolating on $x_{\mathrm{moe}}$, the model captures this general effect with a small GPU profile set instead of profiling every full-cluster layout.

\subsection{Offline Search Algorithm}

Algorithm~\ref{alg-design-placement} jointly searches these choices for the workload's expected input and output lengths.
The generator enumerates layouts $\mathcal{L}$, where each $\ell$ specifies the number of MoE servers, the number of attention servers, the prefill-to-decode server ratio, and the layout of all servers.
\textsc{FitsMemory} first rejects layouts that cannot hold attention and expert weights or sufficient KV caches.

\begin{algorithm}[t]
\caption{Cross-stack placement search}
\label{alg-design-placement}
\begin{algorithmic}[1]
\State $best \gets \bot$
\ForAll{$\ell \in \mathcal{L}$}
    \If{$\neg \Call{FitsMemory}{\ell}$}
        \State \textbf{continue}
    \EndIf
    \ForAll{$q \in \mathcal{Q}_{\ell}$}
        \State $(B_p,\mathit{ok}_p) \gets \Call{MaxFeasibleBatch}{\ell,q,\mathrm{prefill}}$
        \State $(B_d,\mathit{ok}_d) \gets \Call{MaxFeasibleBatch}{\ell,q,\mathrm{decode}}$
        \If{$\mathit{ok}_p \wedge \mathit{ok}_d$}
            \State $g \gets G(\ell,q)$
            \If{$best=\bot \vee g>best.g$}
                \State $best \gets (\ell,q,B_p,B_d,g)$
            \EndIf
        \EndIf
    \EndFor
\EndFor
\State \Return $best$
\end{algorithmic}
\end{algorithm}

\begin{figure*}[t]
    \includegraphics[width=\linewidth]{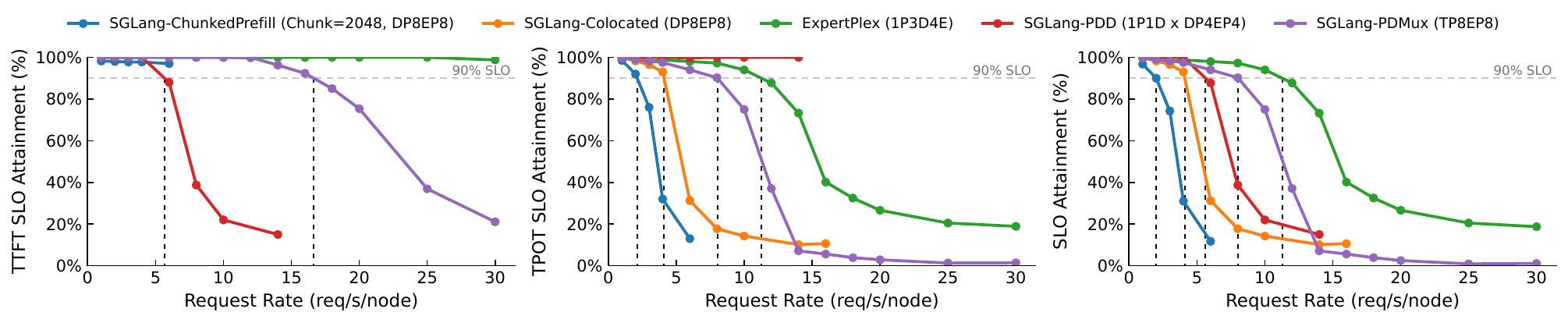}
    \vspace{-0.3in}
    \caption{SLO attainment of \sysname and baselines on the MiniMax-M2.7 and ShareGPT datasets.}
    \label{fig:evaluation:e2e:minimax27_sharegpt_bench_slo_attainment}
\end{figure*}

\begin{figure*}[t]
    \vspace{-0.2in}
    \includegraphics[width=\linewidth]{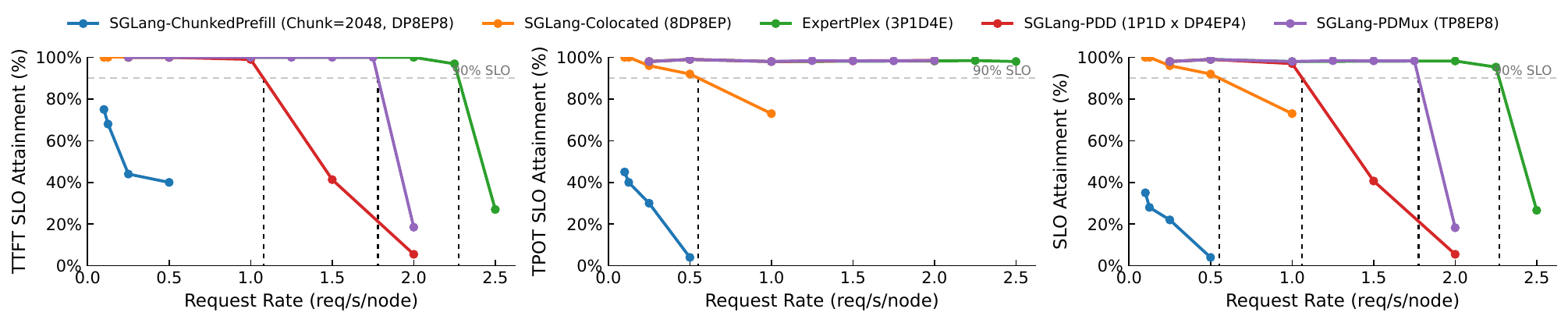}
    \vspace{-0.3in}
    \caption{SLO attainment of \sysname and baselines on the MiniMax-M2.7 and LooGLE datasets.}
    \label{fig:evaluation:e2e:minimax27_loogle_bench_slo_attainment}
\end{figure*}

\begin{figure*}[t]
    \vspace{-0.2in}
    \includegraphics[width=\linewidth]{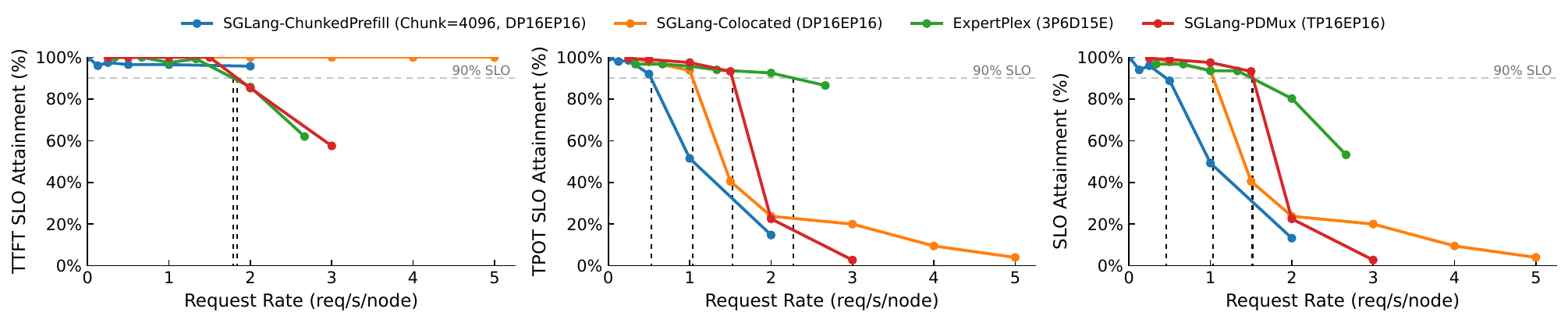}
    \vspace{-0.3in}
    \caption{SLO attainment of \sysname and baselines on the GLM-5.1-FP8 and ShareGPT datasets.}
    \label{fig:evaluation:e2e:glm51_sharegpt_bench_slo_attainment}
\end{figure*}

\begin{figure*}[t]
    \vspace{-0.2in}
    \includegraphics[width=\linewidth]{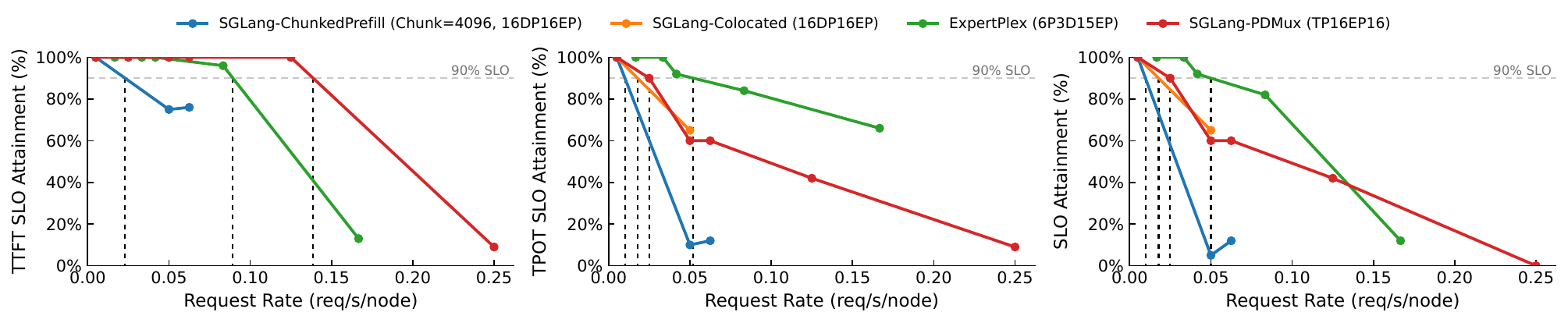}
    \vspace{-0.3in}
    \caption{SLO attainment of \sysname and baselines on the GLM-5.1-FP8 and LooGLE datasets.}
    \label{fig:evaluation:e2e:glm51_loogle_bench_slo_attainment}
    \vspace{-0.2in}
\end{figure*}

For every remaining layout, $\mathcal{Q}_\ell$ enumerates decode SM budgets.
For each $(\ell,q)$, \textsc{MaxFeasibleBatch} binary-searches up to the configured and KV-cache bounds for each phase (lines 6--7).
At each probe, the model from \S\ref{sec:design:model} estimates the attention latency, the tile-aware MoE computation latency under $q$ SMs, and the communication latency.
The overall iteration latency is then estimated as the sum of latencies on the critical path.
Under TBO, \sysname overlaps one microbatch's attention with another's communication and MoE computation, so the iteration latency is the maximum of these two paths.
Under SBO, \sysname overlaps shared-expert computation with routed-expert computation and communication within one microbatch, so the iteration latency is their maximum plus the attention latency.
Lines 8--11 retain only configurations feasible for both phases and select the one with maximum $G$.
The selected configuration jointly fixes the parallelism, placement, overlap strategy, and the expected decode SM budget.

\subsection{Online SM Reallocation}

The offline solution targets an average workload, whereas routing changes $x_{\mathrm{moe}}$ across layers and servers, and either phase may temporarily have no ready MoE computation.
APK therefore interprets $q$ as a contention policy rather than a static partition.
When only one phase is ready, it uses all CTA clusters.
Under contention, it scales $q$ by the ratio of the current decode footprint $x_{\mathrm{moe}}$ to the offline expectation $x_{\mathrm{moe}}^\star$ as follows.
\begin{equation}
q'=\min \left(Q_{\max}, \left\lceil \frac{q x_{\mathrm{moe}}}{x_{\mathrm{moe}}^\star} \right\rceil_c \right),
\end{equation}
Here, $\lceil\cdot\rceil_c$ rounds up to a CTA-cluster multiple, $Q_{\max}$ preserves prefill progress, and prefill receives the remaining clusters.
The reason for protecting decode share first is that decode is more latency-sensitive, and that outputs for already-started requests should be served before inputs for new requests.
For the prefill phase, APK uses all the remaining SMs to preserve throughput as high as possible.

\section{Evaluation}
\label{sec:evaluation}

\subsection{Experiment Setup}

\parabf{Implementation.}
We implement \sysname by modifying DeepGEMM~\cite{deepgemm2025}, DeepEP v1~\cite{deepep}, and SGLang~\cite{sglang}.
The adaptive persistent kernel (APK) is implemented based on DeepGEMM and DeepEP v1.
We fuse two grouped GEMMs, activation functions, MoE data preprocessing, MoE data post-processing and the CTA-cluster scheduler into one long-lived persistent kernel on each MoE GPU.
For each operation, the host determines tile sizes before launch and JIT-compiles them as constants at startup to reduce register pressure.

We also replace DeepEP's receiver-driven all-to-all path with the attention-initiated M-to-N path described in Figure~\ref{fig:overview:architecture} and Figure~\ref{fig:design:comm_overlap}.
Dispatch writes routed activations and metadata directly into the target MoE buffers without receiver-side coordination.
Combine is initiated from the attention side through pull semantics, so the MoE server only publishes finished expert outputs and ready signals.
For the InfiniBand GPUDirect Async (IBGDA) path, we customize \path{nvshmemi_ibgda_get_nbi_warp} to support fine-grained pulls from remote combine buffers over a specific RDMA connection.
In SGLang, we add support for the disaggregated attention and shared-MoE layout, the MoE token dispatcher for this layout, and a \texttt{OperationsStrategy} for communication overlap.

\parabf{Models.}
We use MiniMax-M2.7 for single-node experiments and GLM-5.1-FP8 for multi-node experiments.
Both models use 256 routed experts per MoE layer and activate eight routed experts per token.
MiniMax-M2.7 has a 230~GB FP8 model footprint in our deployment, and each token activates about 7.0B routed expert parameters across layers.
GLM-5.1-FP8 has a 756~GB FP8 model footprint and 724.8B routed expert parameters, of which each token activates about 22.6B.
The former uses full attention; the latter uses DSA.

\begin{figure}[t]
    \centering
    \includegraphics[width=0.98\linewidth]{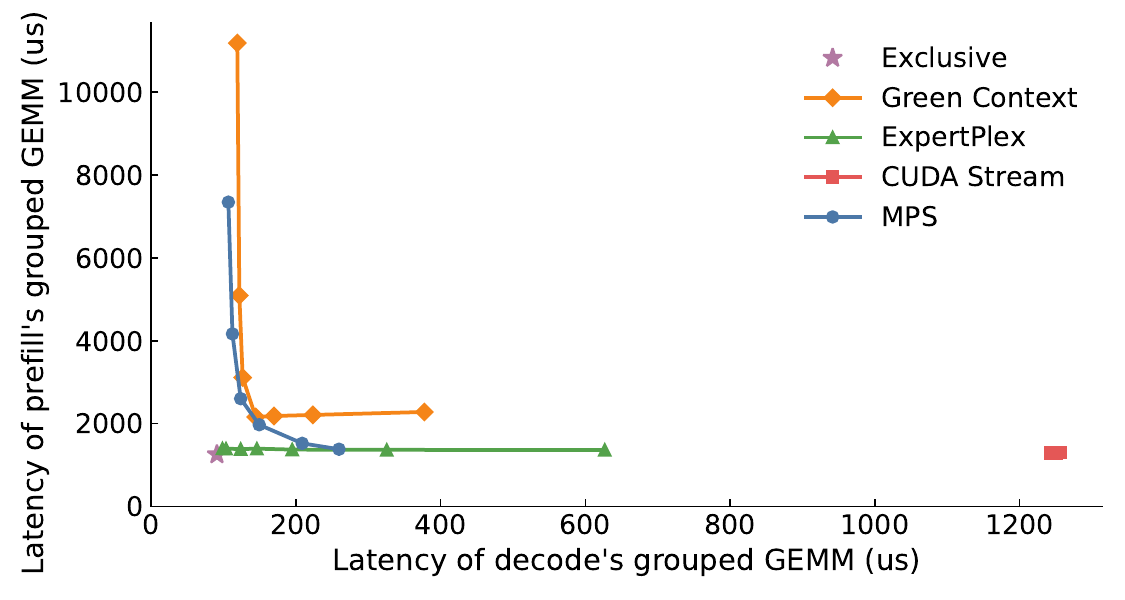}
    \vspace{-0.2in}
    \caption{Pareto frontier of GPU sharing mechanisms.}
    \label{fig:evaluation:preemption_tradeoff}
    \vspace{-0.2in}
\end{figure}

\parabf{Testbed.}
Single-node experiments run on one NVIDIA H800 node with eight GPUs connected by NVLink.
Multi-node experiments use up to three machines.
Each machine has eight NVIDIA H800 GPUs connected by NVLink.
Machines are connected by eight 200~Gbps InfiniBand NICs per node.
All throughput numbers are reported as requests per second per node when systems use different numbers of nodes.

\parabf{Baselines.}
Because \sysname is implemented in SGLang, all baselines use SGLang-based implementations to isolate system-level differences.
\textit{SGLang-Colocated} is the original SGLang serving mode.
It enables DP for attention and EP for MoE.
When memory allows, it enables two-batch overlap to improve throughput.
\textit{SGLang-ChunkedPrefill} additionally enables chunked prefill with the best chunk size found in prior tuning for the corresponding workload and model.
Other settings match SGLang-Colocated.
\textit{SGLang-PDD} implements widely used prefill-decode disaggregation.
Because MoE weights consume most GPU memory, MiniMax-M2.7 uses a 1P1D deployment.
GLM-5.1-FP8 runs out of memory under this PDD layout on 24 GPUs, so we do not report GLM-5.1-FP8 PDD numbers.
\textit{SGLang-PDMux} is based on the open-source MuxWise implementation.
MuxWise targets dense models and uses Green Context to partition GPU resources, so we modify it to support MoE models.
Because its implementation is compatible only with tensor-parallel attention, it uses TP for attention and EP for MoE.
Several baselines cannot express \sysname's 24-GPU fine-grained layout under their parallelism constraints.
For GLM-5.1-FP8, those baselines run on the largest compatible 16-GPU layout, and we compare request rates per node for a fair comparison.

\parabf{Workloads.}
Following prior works, we sample input and output lengths from ShareGPT and LooGLE to represent short and long requests.
Request arrivals follow a Poisson process.
Because SGLang-PDD leaves less GPU memory for the KV cache after duplicating MoE weights across prefill and decode instances, we cap sampled sequence lengths at the PDD KV-cache capacity.

\parabf{Metrics.}
The main metric is P90 goodput, defined as the highest arrival rate at which at least 90\% of requests meet both their time to first token (TTFT) and time per output token (TPOT) SLOs.
For MiniMax-M2.7 on ShareGPT, the TTFT SLO is 1~s and the TPOT SLO is 50~ms.
For MiniMax-M2.7 on LooGLE, the TTFT SLO is 10~s and the TPOT SLO is 100~ms.
For GLM-5.1-FP8 on ShareGPT, the TTFT SLO is 2~s and the TPOT SLO is 100~ms.
For GLM-5.1-FP8 on LooGLE, the TTFT SLO is 20~s and the TPOT SLO is 100~ms.
These SLOs follow the range used by prior disaggregated and high-goodput serving work~\cite{step3afd,megascaleinfer,janus,muxwise}.

\subsection{End-to-End Performance}

Figures~\ref{fig:evaluation:e2e:minimax27_sharegpt_bench_slo_attainment}, \ref{fig:evaluation:e2e:minimax27_loogle_bench_slo_attainment}, \ref{fig:evaluation:e2e:glm51_sharegpt_bench_slo_attainment}, and \ref{fig:evaluation:e2e:glm51_loogle_bench_slo_attainment} show the goodput of all systems on different models and workloads.
Across the four settings, \sysname improves goodput by matching phase resources at the MoE-attention boundary instead of duplicating the whole model or partitioning every GPU by phase.

For MiniMax-M2.7 on ShareGPT, \sysname reaches 11.3 requests per second per node under the joint SLO.
This is 5.65$\times$ higher than SGLang-ChunkedPrefill, 2.72$\times$ higher than SGLang-Colocated, 2.01$\times$ higher than SGLang-PDD, and 1.41$\times$ higher than SGLang-PDMux.

\begin{figure}[t]
    \centering
    \includegraphics[width=0.98\linewidth]{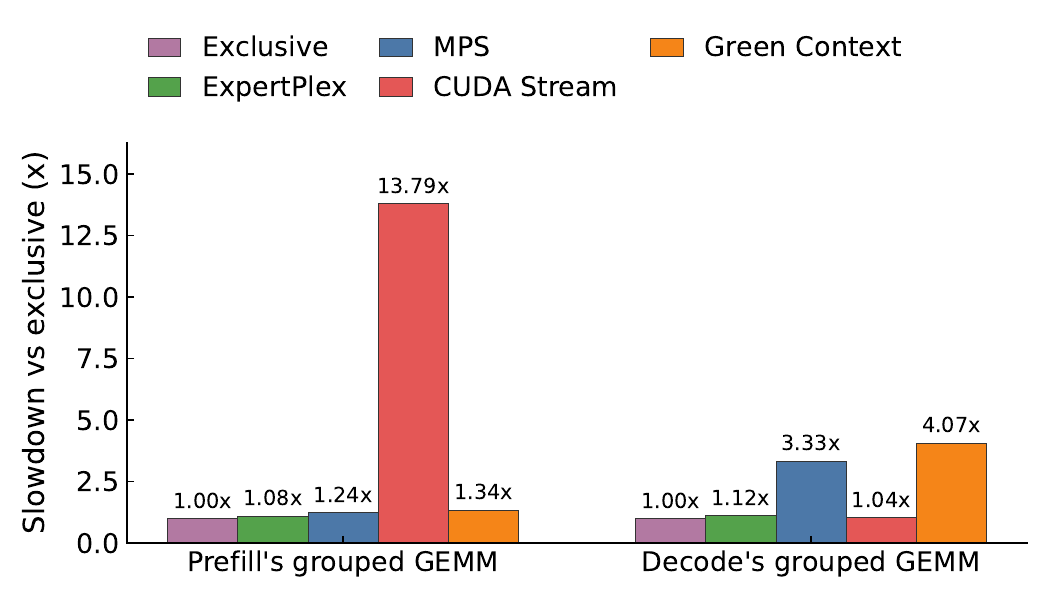}
    \vspace{-0.2in}
    \caption{Case study of preemption slowdown.}
    \label{fig:evaluation:preemption_slowdown}
    \vspace{-0.3in}
\end{figure}

Chunked prefill performs worst.
Each chunk rereads model weights and the KV cache and repeats MoE communication.
Large chunks still interfere with decode, whereas small chunks reduce prefill efficiency.
Colocated serving runs prefill aggressively to build larger batches and improve throughput, but lets long prefill delay decode iterations.
SGLang-PDD removes prefill-decode interference, but it duplicates massive MoE weights and leaves less room for KV cache.
Instance-level disaggregation also leads to resource mismatch, so the prefill phase uses only part of the available resources even when decode is not the bottleneck.
SGLang-PDMux improves over colocated serving by using Green Context, yet its GPU partitions cannot follow the actual number of active experts on each MoE rank.
It also introduces cross-phase communication interference and requires a larger degree of parallelism.
\sysname avoids these failure modes by sharing the MoE pool with APKs while disaggregating attention modules.

On MiniMax-M2.7 with LooGLE, longer requests magnify the difference.
SGLang-ChunkedPrefill cannot sustain the SLO over the evaluated load range because longer inputs amplify its memory-access overhead.
They also amplify interference in SGLang-Colocated, over which \sysname improves goodput by 4.12$\times$; the improvement over SGLang-PDMux is 1.28$\times$ on this workload.

GLM-5.1-FP8 shows similar improvements in multi-node deployments.
Compared with SGLang-ChunkedPrefill, \sysname improves goodput by 3.3$\times$ on ShareGPT and 5.0$\times$ on LooGLE.
Compared with SGLang-Colocated, \sysname improves goodput by 1.5$\times$ on ShareGPT and 2.5$\times$ on LooGLE.
On ShareGPT, \sysname and SGLang-PDMux achieve similar joint goodput of about 1.5 requests per second per node.
PDMux's tensor-parallel attention gives each short request more parallelism and thus a TTFT advantage, but adds communication and allocates resources without regard to MoE sparsity.
On LooGLE, this advantage fades and \sysname improves goodput over SGLang-PDMux by 1.66$\times$.
Long requests expose network interference and Green Context's inability to maximize GPU utilization while preserving TPOT.

\begin{figure}[t]
    \centering
    \includegraphics[width=0.98\linewidth]{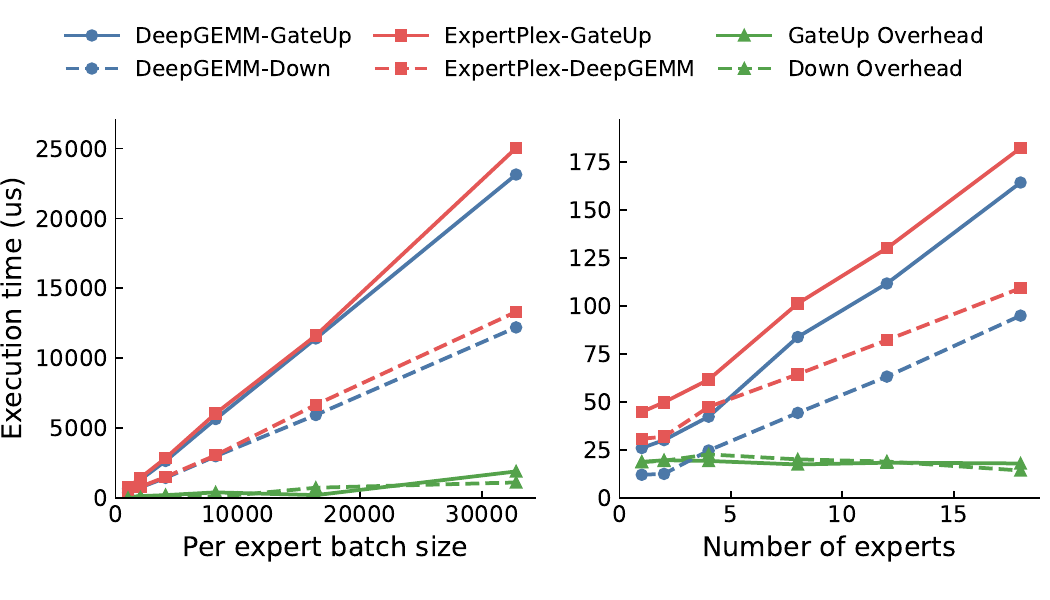}
    \vspace{-0.1in}
    \begin{minipage}{0.98\linewidth}
        \vspace{-0.25in}
        \centering
        \small
        \makebox[0.49\linewidth]{(a) Contiguous layout.}
        \makebox[0.49\linewidth]{(b) Masked layout.}
    \end{minipage}
    \vspace{-0.1in}
    \caption{Overhead of adaptive persistent kernels.}
    \label{fig:evaluation:persistent_grouped_gemm_overhead_glm51_ep15}
    \vspace{-0.2in}
\end{figure}

\subsection{Effectiveness of APKs}
\label{sec:evaluation:apk}

We evaluate APK's GPU sharing mechanism with concurrent prefill- and decode-shaped grouped GEMMs from GLM-5.1-FP8 on one GPU.
The decode grouped GEMM is launched 10~$\mu$s after the prefill grouped GEMM.
The number of tokens is 128 in the decode phase and 8192 in the prefill phase, and all activate eight experts.
We compare APK with exclusive execution, priority CUDA streams, MPS, and Green Context.

Figure~\ref{fig:evaluation:preemption_tradeoff} plots the Pareto frontier.
CUDA stream priorities order kernel launches but provide no isolation after launch, so decode waits behind a running prefill GEMM, leading to significant head-of-line blocking.
Green Context spatially isolates the phases, but its fixed partition prevents either phase from borrowing the other's idle SMs, leading to resource bubbles.
MPS multiplexes space more flexibly, but still cannot time-multiplex an SM at tile granularity within a running kernel.
APK is the only mechanism that reaches the low-latency region for the decode phase while keeping the prefill phase high-performing.

Figure~\ref{fig:evaluation:preemption_slowdown} shows the result under the decode-latency protection target.
CUDA streams increase decode latency by 13.79$\times$ over exclusive execution.
MPS and Green Context keep latency of the decode phase close to exclusive execution, but their fixed allocations slow prefill by 3.33$\times$ and 4.07$\times$.
In contrast, \sysname adds only 8\% overhead to the decode phase and slows prefill by only 1.12$\times$.
The reason is that decode GEMMs are short and intermittent. APK gives all idle CTA clusters to prefill, only preempts prefill when decode arrives, and reallocates idle SMs back to prefill when decode completes the execution.

\begin{figure}[t]
    \centering
    \includegraphics[width=0.98\linewidth]{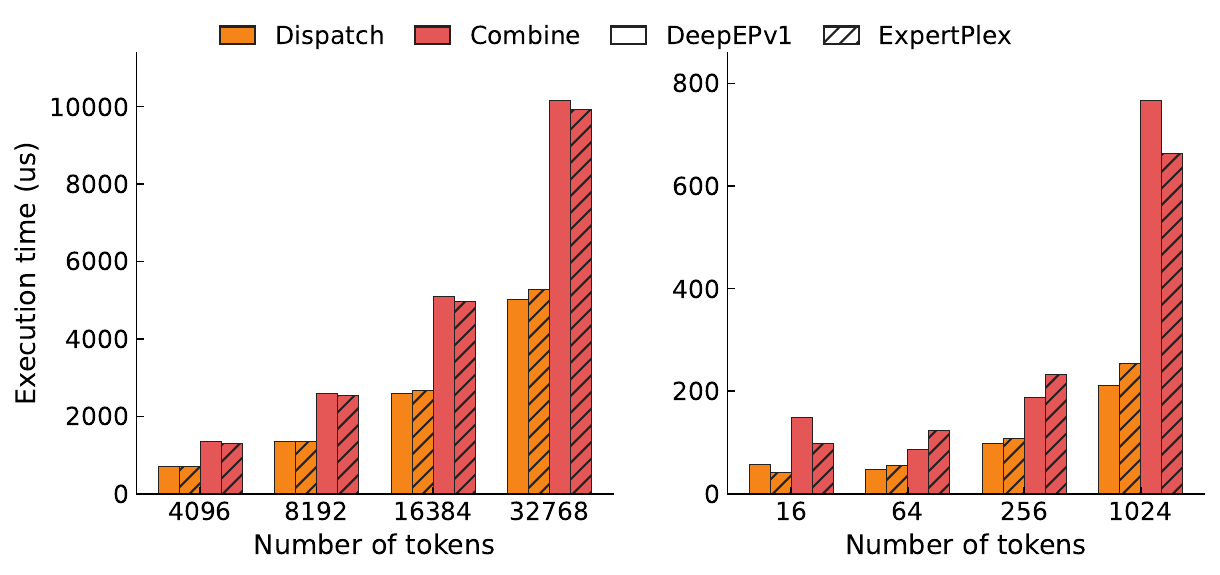}
        \vspace{-0.1in}
    \begin{minipage}{0.98\linewidth}
        \vspace{-0.2in}
        \centering
        \small
        \makebox[0.49\linewidth]{(a) Normal mode.}
        \makebox[0.49\linewidth]{(b) Low-latency mode.}
    \end{minipage}
    \vspace{-0.1in}
    \caption{Overhead of attention-initiated communication.}
    \label{fig:evaluation:internode_glm51_ep8_comm_compare_breakdown}
    \vspace{-0.2in}
\end{figure}

\subsection{Overhead of Tile-Level Scheduling}

Tile-level scheduling adds queue checks, task selection, and CTA-budget enforcement to grouped GEMM.
We measure its overhead against DeepGEMM using the configuration from \S\ref{sec:evaluation:apk}.
Figure~\ref{fig:evaluation:persistent_grouped_gemm_overhead_glm51_ep15}(a) evaluates the contiguous layout used by prefill.
The x-axis varies per-expert batch size, the dominant factor in prefill GEMM time.
Scheduling adds less than 12\% overhead because it runs once per tile group without changing the TMA or Tensor Core pipelines.

Figure~\ref{fig:evaluation:persistent_grouped_gemm_overhead_glm51_ep15}(b) evaluates the masked layout used by decode.
For decode, execution time depends more on the number of active experts than on the number of tokens.
Scheduling adds less than 20~$\mu$s across all measured active-expert counts.
This small cost replaces a much longer wait behind prefill kernels. Moreover, because each token can activate eight experts, its relative overhead falls below 10\% for grouped GEMMs with many active experts.

\subsection{Overhead of Attention-Initiated MoE Communication}

Figure~\ref{fig:evaluation:internode_glm51_ep8_comm_compare_breakdown} measures the attention-initiated MoE communication described in \S\ref{sec:design:comm}.
On 16 GPUs, we compare DeepEP v1 with \sysname's attention-initiated one-sided dispatch and combine under both normal and low-latency execution.

In normal mode, dispatch and combine track DeepEP v1 within about 5\%, preserving hierarchical communication's bandwidth benefit.
Low-latency dispatch and combine expose measurement noise, yet the difference remains within about 45~$\mu$s.
Thus, removing MoE-side coordination and using pull-based combine integrates communication with APK without sacrificing communication efficiency, while leaving MoE-server SMs available to the other phase.

\subsection{Analysis of Preemption Intervals}

\begin{figure}[t]
    \centering
    \includegraphics[width=0.98\linewidth]{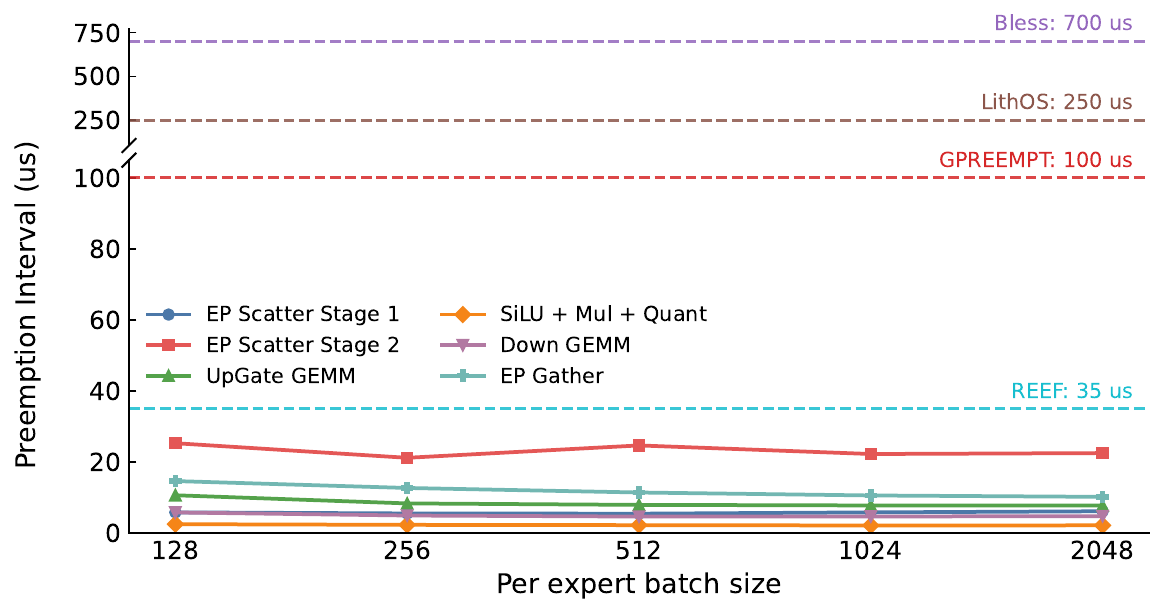}
    \vspace{-0.2in}
    \caption{Preemption interval of different tasks.}
    \label{fig:evaluation:preemption_interval}
    \vspace{-0.2in}
\end{figure}

\sysname tunes GEMM and data-processing tiles for operation performance rather than shrinking them to improve preemption.
Under these tile sizes, Figure~\ref{fig:evaluation:preemption_interval} shows that all MiniMax-M2.7 MoE operations have intervals below 25.3~$\mu$s, and particularly GEMM intervals stay below 10.7~$\mu$s.
The interval depends on tile execution time, not token count, allowing a short decode operation to preempt work from a prefill request with thousands of input tokens.

We compare against the best reported delays of LithOS, Bless, GPreempt, PipeSwitch, and REEF~\cite{bai2020pipeswitch,preemptionInference,lithos,gpreempt,bless} only as reference points, since they do not target MoE LLM workloads and lack the support of modern MoE kernel features such as TMA multicast, CTA clusters, warp specialization, and CUDA Graph.
APK instead schedules at natural tile boundaries, performs useful work throughout each interval, and requires no checkpoint, restore, or recomputation.
Even REEF's best reported delay is 35~$\mu$s and requires recomputing the preempted kernel, which is costly when millisecond-scale prefill GEMMs are frequently interrupted by decode.
APK therefore achieves finer-grained preemption with lower pause and resume costs for MoE serving workloads.

\section{Related Work}
\label{sec:related}

\parabf{LLM serving systems.}
Many techniques are proposed to adapt to accelerate LLM serving.
Prefill-decode disaggregation (PDD) provisions independent prefill and decode instances to avoid interference between two phases~\cite{zhong2024distserve,patel2023splitwise,hu2024inference,mooncake}.
As discussed in \S\ref{sec:background}, instance-level PDD is inefficient to deploy on small clusters, enlarges the rank-failure blast radius and only supports coarse elastic scaling.
Chunked prefill instead splits prefill into smaller chunks to reduce interference~\cite{sarathi,podattention}, while repeated chunk execution introduces extra memory access.
Recent GPU partitioning systems let prefill and decode share a GPU by assigning resources through mechanisms such as Green Context~\cite{muxwise,shi2025nexus,hong2025semi,lin2025bullet}.
However, as described in \S\ref{sec:background}, these systems can introduce bubbles, head-of-line blocking, network interference, and a larger degree of parallelism.
More recent attention-expert disaggregation systems disaggregate attention and expert computation to fit their different characteristics~\cite{megascaleinfer,step3afd,janus}.
However, these designs are built on instance-level PDD, so they inherit the same limitations and further introduce pipeline bubbles~\cite{liu2026revealing}.
In contrast, \sysname shares the MoE servers between phases, which maximizes GPU occupancy and reduces bubbles.
Its attention-initiated communication further introduces communication-computation overlap across phases.
Attention-side techniques such as sequence parallelism~\cite{wu2024loongserve,wu2025tokenlake} and attention offloading~\cite{wang2025prefill,liang2025injecting,lin2024infinitellm} are orthogonal to \sysname.
MoE load-balancing~\cite{he2021fastmoe,he2022fastermoe,lina,smartmoe,wei2026ultraep} solutions are also orthogonal, while \sysname can absorb phase-level load shifts through multiplexing.

\parabf{Kernel scheduling.}
Many systems improve GPU efficiency by scheduling work below the operator boundary.
Megakernel systems break kernel boundaries, fuse small or memory-bound operations, and exploit inter-operator parallelism~\cite{rammer,megakernels,eventtensor,nanoflow,sonicmoe,cheng2026mpk,tilert}.
These techniques target small batch sizes or intra-phase efficiency, but \sysname targets high-goodput MoE serving with concurrent prefill and decode phases, where the scheduler must protect decode SLOs while keeping prefill throughput high.
Other systems overlap computation and communication through fine-grained pipelining or kernel fusion~\cite{comet,flux,liu2026revealing}.
They overlap operations within one phase or one layer pipeline.
\sysname keeps these optimizations compatible, but it can also overlap communication from one phase with computation from the other phase.

GPU sharing and preemption systems provide another line of fine-grained scheduling~\cite{tgs,lithos,preemptionInference,bai2020pipeswitch,gpreempt,mps,nvidia-mig}.
These systems target multitasking across models or applications rather than multiple phases of one MoE model.
They also do not support some modern GPU kernel features such as TMA multicast, CTA clusters, warp specialization, and CUDA Graph.
More importantly, their preemption mechanisms also introduce checkpoint, restore, or recomputation overhead.
\sysname uses tile-level scheduling to avoid these issues to guarantee bounded low preemption and reallocation overhead with temporal and spatial multiplexing support.

\section{Conclusion}
\label{sec:conclusion}

We presented \sysname, a high-goodput MoE LLM serving system that disaggregates attention by phase while sharing massive MoE experts.
\sysname shares MoE weights for memory efficiency, disaggregates attention to reduce attention communication of wide parallelism, uses adaptive persistent kernels to multiplex dynamic sparse MoE computation with performance isolation, and initiates one-sided MoE communication from attention servers to avoid cross-phase interference and overlap communication with computation.
Evaluations on MiniMax-M2.7 and GLM-5.1-FP8 show that \sysname improves goodput by up to 5.65$\times$ over chunked prefill, 2.01$\times$ over instance-level prefill-decode disaggregation, 1.66$\times$ over Green Context-based prefill-decode colocation, and 4.12$\times$ over naive colocated serving systems.

\label{lastpage}


\bibliographystyle{style/ACM-Reference-Format}
\bibliography{paper}

\clearpage
\appendix

\end{document}